\documentclass[amsmath,aps,showpacs,a4paper,10pt]{revtex4}

 \usepackage{epsf}
 \usepackage{graphicx}    

\begin{document}

 \newcommand{\be}[1]{\begin{equation}\label{#1}}
 \newcommand{\ee}{\end{equation}}
 \newcommand{\bea}{\begin{eqnarray}}
 \newcommand{\eea}{\end{eqnarray}}
 \def\disp{\displaystyle}

 \def\gsim{ \lower .75ex \hbox{$\sim$} \llap{\raise .27ex \hbox{$>$}} }
 \def\lsim{ \lower .75ex \hbox{$\sim$} \llap{\raise .27ex \hbox{$<$}} }

 \begin{titlepage}

 \begin{flushright}
 arXiv:1112.2270
 \end{flushright}

 \title{\Large \bf Noether Symmetry in $f(T)$ Theory}

 \author{Hao~Wei\,}
 \email[\,email address:\ ]{haowei@bit.edu.cn}
 \affiliation{School of Physics, Beijing Institute
 of Technology, Beijing 100081, China}

 \author{Xiao-Jiao~Guo\,}
 \affiliation{School of Physics, Beijing Institute
 of Technology, Beijing 100081, China}

 \author{Long-Fei~Wang}
 \affiliation{School of Physics, Beijing Institute
 of Technology, Beijing 100081, China}

 \begin{abstract}\vspace{1cm}
 \centerline{\bf ABSTRACT}\vspace{2mm}
 As is well known, symmetry plays an important role in the
 theoretical physics. In particular, the well-known Noether
 symmetry is an useful tool to select models motivated at a
 fundamental level, and find the exact solution to the given
 Lagrangian. In the present work, we try to consider Noether
 symmetry in $f(T)$ theory. At first, we briefly discuss the
 Lagrangian formalism of $f(T)$ theory. In particular, the
 point-like Lagrangian is explicitly constructed. Based on
 this Lagrangian, the explicit form of $f(T)$ theory and the
 corresponding exact solution are found by requiring Noether
 symmetry. In the resulting $f(T)=\mu T^n$ theory, the universe
 experiences a power-law expansion $a(t)\sim t^{2n/3}$.
 Furthermore, we consider the physical quantities corresponding
 to the exact solution, and find that if $n>3/2$ the expansion
 of our universe can be accelerated without invoking dark
 energy. Also, we test the exact solution of this $f(T)$ theory
 with the latest Union2 Type Ia Supernovae (SNIa) dataset which
 consists of 557 SNIa, and find that it can be well consistent
 with the observational data in fact.
 \end{abstract}


 \maketitle

 \end{titlepage}

 \renewcommand{\baselinestretch}{1.1}


\section{Introduction}\label{sec1}

The current accelerated expansion of our universe~\cite{r1}
 has been one of the most active fields in modern cosmology.
 As is well known, it could be due to an unknown energy
 component (dark energy) or a modification to
 general relativity (modified gravity)~\cite{r1,r2}. Today,
 modified gravity theory has been a competitive alternative
 to the familiar dark energy scenario.

In analogy to the well-known $f(R)$ theory, recently a new
 modified gravity theory, namely the so-called $f(T)$ theory,
 has been proposed to drive the current accelerated expansion
 without invoking dark energy. It is a generalized version
 of the so-called teleparallel gravity originally proposed by
 Einstein~\cite{r3,r4}. In teleparallel gravity,
 the Weitzenb\"ock connection is used, rather than the
 Levi-Civita connection which is used in general relativity.
 Following~\cite{r5,r6}, here we briefly review the key
 points of teleparallel gravity and $f(T)$ theory. In this
 work, we consider a spatially flat Friedmann-Robertson-Walker
 (FRW) universe whose spacetime is described by
 \be{eq1}
 ds^2=-dt^2+a^2(t)d{\bf x}^2,
 \ee
 where $a$ is the scale factor. The orthonormal tetrad
 components $e_i(x^\mu)$ relate to the metric through
 \be{eq2}
 g_{\mu\nu}=\eta_{ij}e_\mu^i e_\nu^j\,,
 \ee
 where Latin $i$, $j$ are indices running over 0, 1, 2, 3 for
 the tangent space of the manifold, and Greek $\mu$,~$\nu$ are
 the coordinate indices on the manifold, also running over 0,
 1, 2, 3. In teleparallel gravity, the gravitational action is
 \be{eq3}
 {\cal S}_T=\int d^4 x\,|e|\,T\,,
 \ee
 where $|e|={\rm det}\,(e_\mu^i)=\sqrt{-g}\,$, and for
 convenience we use the units $16\pi G=\hbar=c=1$ throughout.
 The torsion scalar $T$ is defined by
 \be{eq4}
 T\equiv{S_\rho}^{\mu\nu}\,{T^\rho}_{\mu\nu}\,,
 \ee
 where
 \bea
 {T^\rho}_{\mu\nu} &\equiv &-e^\rho_i\left(\partial_\mu e^i_\nu
 -\partial_\nu e^i_\mu\right)\,,\label{eq5}\\
 {K^{\mu\nu}}_\rho &\equiv &-\frac{1}{2}\left({T^{\mu\nu}}_\rho
 -{T^{\nu\mu}}_\rho-{T_\rho}^{\mu\nu}\right)\,,\label{eq6}\\
 {S_\rho}^{\mu\nu} &\equiv &\frac{1}{2}\left({K^{\mu\nu}}_\rho
 +\delta^\mu_\rho {T^{\theta\nu}}_\theta-
 \delta^\nu_\rho {T^{\theta\mu}}_\theta\right)\,.\label{eq7}
 \eea
 For a spatially flat FRW universe, from Eqs.~(\ref{eq4})
 and~(\ref{eq1}), one has
 \be{eq8}
 T=-6H^2,
 \ee
 where $H\equiv\dot{a}/a$ is the Hubble parameter, and a dot
 denotes a derivative with respect to cosmic time $t$. In
 analogy to the well-known $f(R)$ theory, one can replace $T$
 in the gravitational action~(\ref{eq3}) by any function
 $f(T)$, and then obtain the so-called $f(T)$ theory. In $f(T)$
 theory, the modified Friedmann equation and Raychaudhuri
 equation are given by~\cite{r5,r6,r7,r8}
 \bea
 &&12H^2 f_T+f=\rho\,,\label{eq9}\\
 &&48H^2 f_{TT}\dot{H}-f_T\left(12H^2+4\dot{H}\right)-f
 =p\,,\label{eq10}
 \eea
 where a subscript $T$ denotes a derivative with respect
 to $T$, and $\rho$, $p$ are the total energy density and
 pressure, respectively. In an universe with only pressureless
 matter, obviously we have $p=p_m=0$ and
 $\rho=\rho_m=\rho_{m0}\,a^{-3}$, where the subscript ``0''
 indicates the present value of the corresponding quantity, and
 we have set $a_0=1$. It is well known that when $f(T)=T$ the
 familiar general relativity can be completely recovered.

In fact, $f(T)$ theory was firstly used to drive inflation
 by Ferraro and Fiorini~\cite{r9}. Later, Bengochea
 and Ferraro~\cite{r5}, as well as Linder~\cite{r6},
 proposed to use $f(T)$ theory to drive the current accelerated
 expansion of our universe without invoking dark energy. Very
 soon, $f(T)$ theory attracted much attention in the community.
 We refer to e.g.~\cite{r7,r8,r10,r11,r12,r13,r14,r15,r41} for
 relevant works.

So far, the specified forms of function $f(T)$ in
 the literature are written by hand. There is no natural
 guidance from fundamental physics on the form of $f(T)$ in
 fact. In the present work, we try to address this issue. As
 is well known, symmetry plays an important role in the
 theoretical physics. In particular, the well-known Noether
 symmetry is an useful tool to select models motivated at a
 fundamental level. In the literature, Noether symmetry
 has been extensively used in scalar field
 cosmology~\cite{r16,r17}, non-minimally coupled
 cosmology~\cite{r18}, $f(R)$ theory~\cite{r19,r20,r21},
 scalar-tensor theory~\cite{r22}, higher order gravity
 theory~\cite{r23}, multiple scalar fields~\cite{r24},
 vector field~\cite{r25}, fermion field~\cite{r26}, tachyon
 field~\cite{r27}, non-flat cosmology~\cite{r28}, quantum
 cosmology~\cite{r29}, Bianchi universe~\cite{r30}, Brans-Dicke
 theory~\cite{r31}, dilaton~\cite{r32}, induced gravity
 theory~\cite{r33}, gravity with variable $G$ and
 $\Lambda$~\cite{r34}, Gauss-Bonnet gravity~\cite{r35},
 and so on. Now, we try to consider Noether symmetry in $f(T)$
 theory in the present work.

This paper is organized as follows. In Sec.~\ref{sec2}, we
 briefly discuss the Lagrangian formalism of $f(T)$ theory.
 In particular, the point-like Lagrangian is explicitly
 constructed. In Sec.~\ref{sec3}, we consider Noether symmetry
 in $f(T)$ theory. The explicit form of $f(T)$ theory and the
 corresponding exact solution are found by requiring Noether
 symmetry. In Sec.~\ref{sec4}, we further discuss the physical
 quantities corresponding to the exact solution. In
 Sec.~\ref{sec5}, we test the exact solution of $f(T)$ theory
 found in Sec.~\ref{sec3} with the latest Union2 Type Ia
 Supernovae (SNIa) dataset which consists of 557 SNIa. Finally,
 some brief concluding remarks are given in Sec.~\ref{sec6}.


\section{Lagrangian formalism of $f(T)$ theory}\label{sec2}

In the study of Noether symmetry, the point-like Lagrangian
 plays an important role. In this section, we discuss the
 Lagrangian formalism of $f(T)$ theory. As mentioned above,
 the relevant action of $f(T)$ theory is given by
 \be{eq11}
 {\cal S}=\int d^4 x\,|e|\,f(T)+{\cal S}_m\,,
 \ee
 where ${\cal S}_m$ is the action of pressureless matter
 minimally coupled with gravity, and we assume that the
 radiation can be ignored. Following e.g.~\cite{r20,r21},
 to derive the cosmological equations in the FRW metric, one
 can define a canonical Lagrangian
 ${\cal L}={\cal L}(a,\dot{a},T,\dot{T})$, whereas
 ${\cal Q}=\{a,T\}$ is the configuration space, and
 ${\cal TQ}=\{a,\dot{a},T,\dot{T}\}$ is the related tangent
 bundle on which $\cal L$ is defined. The scale factor $a(t)$
 and the torsion scalar $T(t)$ are taken as independent
 dynamical variables. One can use the method of Lagrange
 mutipliers to set $T$ as a constraint of the dynamics (nb.
 Eq.~(\ref{eq8})). Selecting the suitable Lagrange mutiplier
 and integrating by parts, the Lagrangian $\cal L$ becomes
 canonical~\cite{r20,r21}. In our case, we have
 \be{eq12}
 {\cal S}=2\pi^2\int dt\,a^3\left[f(T)-\lambda\left(T+
 6\frac{\dot{a}^2}{a^2}\right)-\frac{\rho_{m0}}{a^3}\right],
 \ee
 where $\lambda$ is a Lagrange mutiplier. The variation with
 respect to $T$ of this action gives
 \be{eq13}
 \lambda=f_T\,.
 \ee
 Therefore, the action~(\ref{eq12}) can be rewritten as
 \be{eq14}
 {\cal S}=2\pi^2\int dt\,a^3\left[f(T)-f_T\left(T+
 6\frac{\dot{a}^2}{a^2}\right)-\frac{\rho_{m0}}{a^3}\right],
 \ee
 and then the point-like Lagrangian reads (up to a constant
 factor $2\pi^2$)
 \be{eq15}
 {\cal L}(a,\dot{a},T,\dot{T})=a^3\left(f-f_T T\right)-
 6f_T a\dot{a}^2-\rho_{m0}\,.
 \ee
 As is well known, for a dynamical system, the Euler-Lagrange
 equation is
 \be{eq16}
 \frac{d}{dt}\left(\frac{\partial \cal L}{\partial\dot{q}_i}
 \right)-\frac{\partial \cal L}{\partial q_i}=0\,,
 \ee
 where $q_i$ are the generalized coordinates of
 the configuration space $\cal Q$, and in our case $q_i=a$
 and $T$. Substituting Eq.~(\ref{eq15}) into the Euler-Lagrange
 equation~(\ref{eq16}), we obtain
 \bea
 &&a^3 f_{TT}\left(T+
 6\frac{\dot{a}^2}{a^2}\right)=0\,,\label{eq17}\\
 &&f-f_T T+2f_T H^2+
 4\left(f_T\frac{\ddot{a}}{a}+Hf_{TT}\dot{T}\right)=0\,.\label{eq18}
 \eea
 If $f_{TT}\not=0$, from Eq.~(\ref{eq17}), it is easy to find that
 \be{eq19}
 T=-6\frac{\dot{a}^2}{a^2}=-6H^2\,,
 \ee
 i.e., the relation~(\ref{eq8}) is recovered. Generally, this
 is the Euler constraint of the dynamics. Substituting
 Eq.~(\ref{eq19}) into Eq.~(\ref{eq18}) and using
 $\ddot{a}/a=H^2+\dot{H}$, we get
 \be{eq20}
 48H^2 f_{TT}\dot{H}-4f_T\left(3H^2+\dot{H}\right)-f=0\,,
 \ee
 i.e., the modified Raychaudhuri equation~(\ref{eq10}) is
 recovered (note that $p=p_m=0$). On the other hand, it is also
 well known that the total energy (Hamiltonian) corresponding
 to Lagrangian $\cal L$ is given by
 \be{eq21}
 {\cal H}=\sum\limits_i\frac{\partial \cal L}
 {\partial\dot{q}_i}\dot{q}_i-{\cal L}\,.
 \ee
 Substituting Eq.~(\ref{eq15}) into Eq.~(\ref{eq21}), we have
 \be{eq22}
 {\cal H}(a,\dot{a},T,\dot{T})=a^3\left(-6f_T
 \frac{\dot{a}^2}{a^2}-f+f_T T+\frac{\rho_{m0}}{a^3}\right).
 \ee
 Considering the total energy ${\cal H}=0$ (Hamiltonian
 constraint)~\cite{r16,r20,r21} and using Eq.~(\ref{eq19}),
 we get
 \be{eq23}
 12H^2 f_T+f=\frac{\rho_{m0}}{a^3}\,,
 \ee
 i.e., the modified Friedmann equation~(\ref{eq9}) is also
 recovered (note that $\rho=\rho_m=\rho_{m0}/a^3$). So far, we
 have shown that the point-like Lagrangian given
 in Eq.~(\ref{eq15}) can yield all the correct equations of
 motion, and hence it is the desired one.


\section{Noether symmetry in $f(T)$ theory}\label{sec3}

As is well known (see e.g.~\cite{r16,r17,r18,r19,r20,r21,r22,
r23,r24,r25,r26,r27,r28,r29,r30,r31,r32,r33,r34,r35}), Noether
 symmetry is an useful tool to select models motivated at a
 fundamental level, and find the exact solution to the given
 Lagrangian. In this section, we try to consider Noether
 symmetry in $f(T)$ theory.

Following e.g.~\cite{r20,r21}, the generator of Noether
 symmetry is a vector
 \be{eq24}
 {\bf X}=\alpha\frac{\partial}{\partial a}+
 \beta\frac{\partial}{\partial T}+
 \dot{\alpha}\frac{\partial}{\partial\dot{a}}+
 \dot{\beta}\frac{\partial}{\partial\dot{T}}\,,
 \ee
 where $\alpha=\alpha(a,T)$ and $\beta=\beta(a,T)$ are both
 functions of the generalized coordinates $a$ and $T$. Noether
 symmetry exists if the equation
 \be{eq25}
 L_{\bf X}{\cal L}={\bf X}{\cal L}=
 \alpha\frac{\partial \cal L}{\partial a}+
 \beta\frac{\partial \cal L}{\partial T}+
 \dot{\alpha}\frac{\partial \cal L}{\partial\dot{a}}+
 \dot{\beta}\frac{\partial \cal L}{\partial\dot{T}}=0
 \ee
 has solution, where $L_{\bf X}{\cal L}$ is the Lie derivative
 of the Lagrangian $\cal L$ with respect to the vector $\bf X$.
 Of course, according to the well-known Noether theorem, there
 will be a constant of motion (Noether charge)~\cite{r20,r21},
 namely
 \be{eq26}
 Q_0=\sum\limits_i\alpha_i\frac{\partial \cal L}{\partial\dot{q}_i}
 =\alpha\frac{\partial \cal L}{\partial\dot{a}}+
 \beta\frac{\partial \cal L}{\partial\dot{T}}=const.
 \ee
 Note that the meaning of $L_{\bf X}{\cal L}=0$ is that $\cal L$ is
 constant along the flow (possibly a local flow) generated by
 $\bf X$, namely Eq.~(\ref{eq25}) is identically verified all
 over $\cal TQ$~\cite{r16}. Its explicit evaluation gives an
 expression of second degree in $\dot{a}$ and $\dot{T}$,
 whose coefficients are functions of $a$ and $T$ only.
 Therefore, they should be zero separately~\cite{r16,r20,r21}.

In our case, substituting Eq.~(\ref{eq15}) into
 Eq.~(\ref{eq25}) and using the relations $\dot{\alpha}=
 (\partial\alpha/\partial a)\,\dot{a}
 +(\partial\alpha/\partial T)\,\dot{T}$, $\dot{\beta}=
 (\partial\beta/\partial a)\,\dot{a}+
 (\partial\beta/\partial T)\,\dot{T}$, we obtain
 \be{eq27}
 3\alpha a^2\left(f-f_T T\right)-\beta a^3 f_{TT}T
 -6\dot{a}^2\left(\alpha f_T+\beta a f_{TT}+
 2a f_T\frac{\partial\alpha}{\partial a}\right)-
 12a\dot{a}\dot{T}\frac{\partial\alpha}{\partial T}=0\,.
 \ee
 As mentioned above, requiring the coefficients of $\dot{a}^2$,
 $\dot{T}^2$ and $\dot{a}\dot{T}$ in Eq.~(\ref{eq27}) to be
 zero, we find that
 \bea
 &&a\frac{\partial\alpha}{\partial T}=0\,,\label{eq28}\\
 &&\alpha f_T+\beta a f_{TT}+
 2a f_T\frac{\partial\alpha}{\partial a}=0\,,\label{eq29}\\
 &&3\alpha a^2\left(f-f_T T\right)-
 \beta a^3 f_{TT}T=0\,.\label{eq30}
 \eea
 In particular, the constraint~(\ref{eq30}) is sometimes
 called Noether condition~\cite{r20}. The corresponding
 constant of motion (Noether charge) given in Eq.~(\ref{eq26})
 reads
 \be{eq31}
 Q_0=-12\alpha f_T a\dot{a}=const.
 \ee
 A solution of Eqs.~(\ref{eq28}), (\ref{eq29}) and~(\ref{eq30})
 exists if explicit forms of $\alpha$ and $\beta$ are found;
 and if at least one of them is different from zero, Noether
 symmetry exists~\cite{r20}. Obviously, from Eq.~(\ref{eq28}),
 it is easy to see that $\alpha$ is independent of $T$, and
 hence it is a function of $a$ only, i.e., $\alpha=\alpha(a)$.
 On the other hand, from Eq.~(\ref{eq30}), we have
 \be{eq32}
 \beta a f_{TT}T=3\alpha\left(f-f_T T\right).
 \ee
 Multiplying $T$ in both sides of Eq.~(\ref{eq29}), and then
 substituting Eq.~(\ref{eq32}) into it, we obtain
 \be{eq33}
 f_T T\left(2a\frac{d\alpha}{da}-2\alpha\right)+3\alpha f=0\,.
 \ee
 Fortunately, one can perform a separation of variables, and
 recast Eq.~(\ref{eq33}) as
 \be{eq34}
 1-\frac{a}{\alpha}\frac{d\alpha}{da}=\frac{3f}{2f_T T}\,.
 \ee
 Since its left-hand side is a function of $a$ only and its
 right-hand side is a function of $T$ only, they must be
 equal to a same constant in order to ensure that
 Eq.~(\ref{eq34}) always holds. For convenience, we let this
 constant be $3/(2n)$, and then Eq.~(\ref{eq34}) can be
 separated into two ordinary differential equations, i.e.,
 \bea
 &&nf=f_T T\,,\label{eq35}\\
 &&1-\frac{a}{\alpha}\frac{d\alpha}{da}=\frac{3}{2n}\,.\label{eq36}
 \eea
 It is easy to find the solutions of these two ordinary
 differential equations, namely
 \bea
 &&f(T)=\mu T^n\,,\label{eq37}\\
 &&\alpha(a)=\alpha_0\,a^{1-3/(2n)}\,,\label{eq38}
 \eea
 where $\mu$ and $\alpha_0$ are integral constants. Obviously,
 $f(T)$ and $\alpha(a)$ are both power-law forms. Substituting
 Eqs.~(\ref{eq37}) and~(\ref{eq38}) into Eq.~(\ref{eq32}), we
 find that
 \be{eq39}
 \beta(a,T)
 =-\frac{3\alpha_0}{n}\,a^{-3/(2n)}\,T\,.
 \ee
 So far, we have found the explicit non-zero solutions of
 $f(T)$, $\alpha$ and $\beta$. Therefore, Noether symmetry
 exists. Finally, we try to find out the exact solution of
 $a(t)$ for this type of $f(T)$. Substituting
 Eqs.~(\ref{eq37}), (\ref{eq38}) and~(\ref{eq19}) into
 Eq.~(\ref{eq31}), we can obtain an ordinary differential
 equation of $a(t)$, namely
 \be{eq40}
 a^{c_1}\,\dot{a}=c_2\,,
 \ee
 where
 \be{eq41}
 c_1=\frac{3}{2n}-1\,,~~~~~~~
 c_2=\left[\frac{Q_0}{-12\alpha_0\mu n(-6)^{n-1}}
 \right]^{1/(2n-1)}\,.
 \ee
 It is easy to find that the solution of Eq.~(\ref{eq40}) is
 given by
 \be{eq42}
 a(t)=-(1+c_1)(c_3-c_2 t)^{1/(1+c_1)}
 =(-1)^{1+2n/3}\cdot\frac{3}{2n}\,(c_2 t-c_3)^{2n/3}\,,
 \ee
 where $c_3$ is an integral constant. Obviously, in the late
 time $|c_2 t|\gg |c_3|$ and the universe experiences a
 power-law expansion. In fact, we can make it clearer.
 Requiring $a(t=0)=0$, it is easy to see that the integral
 constant $c_3$ is zero in fact. So, we have
 \be{eq43}
 a(t)\sim t^{2n/3}\,,
 \ee
 where its prefactor
 $\disp (-1)^{1+2n/3}\cdot\frac{3}{2n}\,c_2^{2n/3}$ is not
 important. Note that $n>0$ is required to ensure that the
 universe is expanding.


\section{Physical quantities corresponding to the exact solution}\label{sec4}

Here, we further consider the physical quantities
 corresponding to the exact solution found in the previous
 section. Firstly, from Eq.~(\ref{eq43}), we find that the
 Hubble parameter is
 \be{eq44}
 H\equiv\frac{\dot{a}}{a}=\frac{2n}{3}\,t^{-1}\,,
 \ee
 and the deceleration parameter is given by
 \be{eq45}
 q\equiv-\frac{\ddot{a}}{aH^2}=\frac{3}{2n}-1\,.
 \ee
 When $n>3/2$, the expansion of our universe can
 be accelerated (note that $n<0$ is not acceptable since the
 universe contracts in this case). Secondly, from
 Eqs.~(\ref{eq9}) and~(\ref{eq10}), one can find that the
 effective dark energy density and pressure from torsion are
 given by~\cite{r5,r6,r7,r8}
 \bea
 &&\rho_{de}=6H^2-f-12H^2 f_T\,,\label{eq46}\\
 &&p_{de}=-\rho_{de}-4\left(12H^2 f_{TT}-f_T+
 1\right)\dot{H}\,.\label{eq47}
 \eea
 Substituting Eqs.~(\ref{eq44}), (\ref{eq19}) and~(\ref{eq37})
 into Eqs.~(\ref{eq46}) and (\ref{eq47}), we can find that
 the equation-of-state parameter (EoS) of the effective dark
 energy from torsion is given by
 \be{eq48}
 w_{de}\equiv\frac{p_{de}}{\rho_{de}}
 =-\frac{8n(n-1)\,3^n}{8n^2\cdot 3^n-3\mu(8^n
 -n\cdot 2^{1+3n})(-n^2)^n\,t^{2(1-n)}}\,.
 \ee
 Using the well-known relation between the total EoS $w_{tot}$
 and the deceleration parameter $q$ (see e.g.~\cite{r36}), it
 is easy to see that
 \be{eq49}
 w_{tot}\equiv\frac{p_{tot}}{\rho_{tot}}=
 \frac{1}{3}\left(2q-1\right)=\frac{1}{n}-1\,.
 \ee
 Again, when $n>3/2$ we have $w_{tot}<-1/3$ and then the
 expansion of our universe can be accelerated (note that $n<0$
 is not acceptable since the universe contracts in this case).
 It is worth noting that if $n>1$, from Eq.~(\ref{eq48}) we
 have $w_{de}\to 1/n-1=w_{tot}$ in the late time
 $t\to\infty$. This is not surprising. Note that the
 well-known relation (see e.g.~\cite{r36})
 \be{eq50}
 w_{tot}=\Omega_{de}w_{de}+\Omega_m w_m\,,
 \ee
 where the EoS of pressureless matter $w_m=0$, and $\Omega_i$
 are the fractional energy densities of the effective dark
 energy and pressureless matter. Although $w_{tot}$ is
 constant, $w_{de}$ and $\Omega_{de}$ are dependent on time
 $t$. If $w_{de}<w_m=0$, it is inevitable that
 $\Omega_{de}\to 1$ and then $w_{de}\to w_{tot}$ in the late
 time $t\to\infty$. Finally, we turn to the fractional energy
 densities. Using Eqs.~(\ref{eq48}), (\ref{eq49}), (\ref{eq50})
 and $w_m=0$, we find that the fractional density of the
 effective dark energy from torsion is given by
 \be{eq51}
 \Omega_{de}=\frac{w_{tot}}{w_{de}}=
 1-\frac{3\mu(8^n-n\cdot 2^{1+3n})(-n^2)^n\,
 t^{2(1-n)}}{8n^2\cdot 3^n}\,,
 \ee
 and then the fractional density of pressureless matter
 $\Omega_m=1-\Omega_{de}$ is ready, namely
 \be{eq52}
 \Omega_m=\frac{3\mu(8^n-n\cdot 2^{1+3n})(-n^2)^n
 \,t^{2(1-n)}}{8n^2\cdot 3^n}\,.
 \ee
 It is worth noting that $\Omega_m$ coming from
 Eq.~(\ref{eq52}) contains only two parameters $n$ and $\mu$.
 On the other hand, by definition $\Omega_m\propto\rho_m/H^2$
 and $\rho_m=\rho_{m0}/a^3$, while the prefactor in $a(t)$
 contains $c_2$ which is given in Eq.~(\ref{eq41}), we see that
 $\Omega_m$ contains five parameters $\rho_{m0}$, $n$, $\mu$,
 $\alpha_0$ and $Q_0$ in this way. Requiring the equality
 between these two $\Omega_m$, it is easy to see that at least
 one of these five parameters $\rho_{m0}$, $n$, $\mu$,
 $\alpha_0$ and $Q_0$ is not independent.


\section{Cosmological test}\label{sec5}

As is well known, the current accelerated expansion of our
 universe was firstly found from the observation of distant
 Type Ia Supernovae (SNIa)~\cite{r1}. Here, we would like
 to test the exact solution of $f(T)$ theory found in
 Sec.~\ref{sec3} with the latest Union2 SNIa
 dataset~\cite{r37} which consists of 557 SNIa.

The data points of the 557 Union2 SNIa compiled
 in~\cite{r37} are given in terms of the distance modulus
 $\mu_{obs}(z_i)$. On the other hand, the theoretical
 distance modulus is defined as
 \be{eq53}
 \mu_{th}(z_i)\equiv 5\log_{10}D_L(z_i)+\mu_0\,,
 \ee
 where $z=1/a-1$ is redshift, $\mu_0\equiv 42.38-5\log_{10}h$
 and $h$ is the Hubble constant $H_0$ in units of
 $100~{\rm km/s/Mpc}$, whereas
 \be{eq54}
 D_L(z)=(1+z)\int_0^z \frac{d\tilde{z}}{E(\tilde{z};{\bf p})}\,,
 \ee
 in which $E\equiv H/H_0$, and ${\bf p}$ denotes the model
 parameters. Correspondingly, the $\chi^2$ from the 557
 Union2 SNIa is given by
 \be{eq55}
 \chi^2_{\mu}({\bf p})=\sum\limits_{i}\frac{\left[
 \mu_{obs}(z_i)-\mu_{th}(z_i)\right]^2}{\sigma^2(z_i)}\,,
 \ee
 where $\sigma$ is the corresponding $1\sigma$ error. The parameter
 $\mu_0$ is a nuisance parameter but it is independent of the data
 points. One can perform an uniform marginalization over $\mu_0$.
 However, there is an alternative way. Following~\cite{r38,r39}, the
 minimization with respect to $\mu_0$ can be made by expanding the
 $\chi^2_{\mu}$ of Eq.~(\ref{eq55}) with respect to $\mu_0$ as
 \be{eq56}
 \chi^2_{\mu}({\bf p})=\tilde{A}-2\mu_0\tilde{B}+\mu_0^2\tilde{C}\,,
 \ee
 where
 $$\tilde{A}({\bf p})=\sum\limits_{i}\frac{\left[\mu_{obs}(z_i)
 -\mu_{th}(z_i;\mu_0=0,{\bf p})\right]^2}
 {\sigma_{\mu_{obs}}^2(z_i)}\,,$$
 $$\tilde{B}({\bf p})=\sum\limits_{i}\frac{\mu_{obs}(z_i)
 -\mu_{th}(z_i;\mu_0=0,{\bf p})}{\sigma_{\mu_{obs}}^2(z_i)}\,,
 ~~~~~~~~~~
 \tilde{C}=\sum\limits_{i}\frac{1}{\sigma_{\mu_{obs}}^2(z_i)}\,.$$
 Eq.~(\ref{eq56}) has a minimum for
 $\mu_0=\tilde{B}/\tilde{C}$ at
 \be{eq57}
 \tilde{\chi}^2_{\mu}({\bf p})=
 \tilde{A}({\bf p})-\frac{\tilde{B}({\bf p})^2}{\tilde{C}}\,.
 \ee
 Since $\chi^2_{\mu,\,min}=\tilde{\chi}^2_{\mu,\,min}$ obviously (up
 to a constant), we can instead minimize $\tilde{\chi}^2_{\mu}$
 which is independent of $\mu_0$. As is well known, the
 best-fit model parameters are determined by minimizing
 $\chi^2=\tilde{\chi}^2_{\mu}$. As in~\cite{r38,r40}, the
 $68.3\%$ confidence level is determined by
 $\Delta\chi^2\equiv\chi^2-\chi^2_{min}\leq 1.0$, $2.3$ and
 $3.53$ for $n_p=1$, $2$ and $3$, respectively, where $n_p$ is
 the number of free model parameters. Similarly, the $95.4\%$
 confidence level is determined by
 $\Delta\chi^2\equiv\chi^2-\chi^2_{min}\leq 4.0$, $6.17$ and
 $8.02$ for $n_p=1$, $2$ and $3$, respectively. Note that the
 corresponding $h$ can be determined by $\mu_0=\tilde{B}/\tilde{C}$
 for the best-fit parameters.


 \begin{center}
 \begin{figure}[tbhp]
 \centering
 \includegraphics[width=0.98\textwidth]{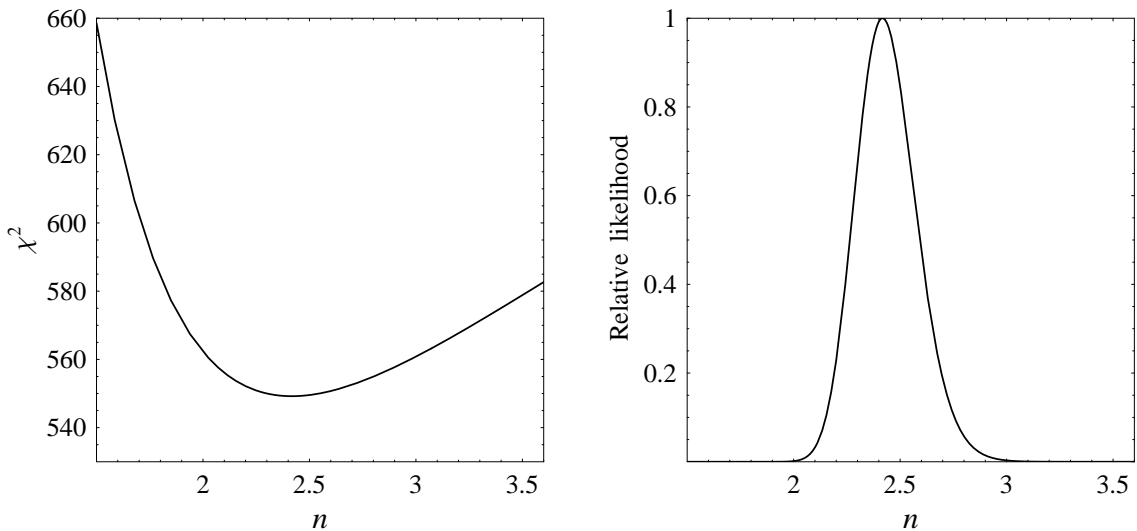}
 \caption{\label{fig1} The $\chi^2$ and likelihood
 ${\cal L}_{\chi^2}\propto e^{-\chi^2/2}$ as
 functions of parameter $n$.}
 \end{figure}
 \end{center}


\vspace{-4mm}  

\begin{center}
 \begin{figure}[tbhp]
 \centering
 \includegraphics[width=0.49\textwidth]{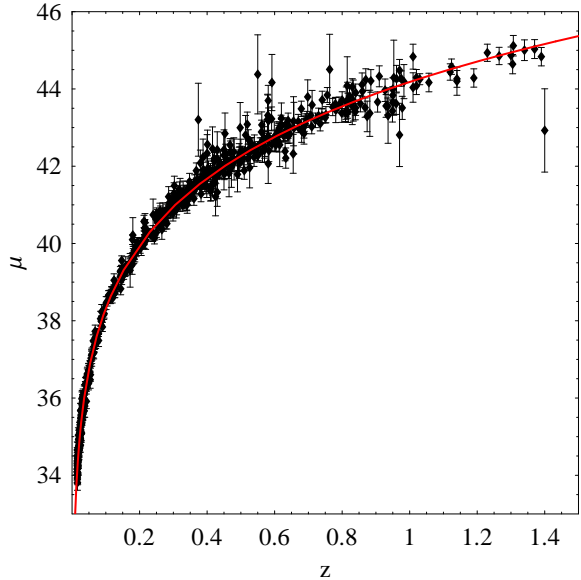}
 \caption{\label{fig2} The Hubble diagram for the best fit
 (red solid line), comparing with the 557 Union2 SNIa data
 points (black diamonds). See the text for details.}
 \end{figure}
 \end{center}

In our case, from Eqs.~(\ref{eq43}) and~(\ref{eq44}), we have
 $H\propto a^{-3/(2n)}$, and then
 \be{eq58}
 E=\frac{H}{H_0}=a^{-3/(2n)}=(1+z)^{3/(2n)}\,.
 \ee
 We present the corresponding $\chi^2$ and
 likelihood ${\cal L}_{\chi^2}\propto e^{-\chi^2/2}$ as
 functions of parameter $n$ in Fig.~\ref{fig1}. The best fit
 has $\chi^2_{min}=549.222$, and the best-fit parameter is
 \be{eq59}
 n=2.417\,^{+0.147}_{-0.131}\ {\rm (with\ 1\sigma
 \ uncertainty)}\ ^{+0.312}_{-0.250}\ {\rm (with\ 2\sigma
 \ uncertainty)}.
 \ee
 The corresponding $h=0.692$ for the best fit. Obviously,
 $n>3/2$ as required by the accelerated expansion of our
 universe. In Fig.~\ref{fig2}, we show the Hubble diagram for
 the best fit, comparing with the 557 Union2 SNIa data points.
 Obviously, our $f(T)$ theory with solution~(\ref{eq43}) can
 be well consistent with the latest 557 Union2 SNIa dataset.


\vspace{-12mm}  


\section{Concluding remarks}\label{sec6}

As is well known, symmetry plays an important role in the
 theoretical physics. In particular, the well-known Noether
 symmetry is an useful tool to select models motivated at a
 fundamental level, and find the exact solution to the given
 Lagrangian. In the present work, we try to consider Noether
 symmetry in $f(T)$ theory. At first, we briefly discuss the
 Lagrangian formalism of $f(T)$ theory. In particular, the
 point-like Lagrangian is explicitly constructed. Based on
 this Lagrangian, the explicit form of $f(T)$ theory and the
 corresponding exact solution are found by requiring Noether
 symmetry. In the resulting $f(T)=\mu T^n$ theory, the
 universe experiences a power-law expansion
 $a(t)\sim t^{2n/3}$. Furthermore, we consider the physical
 quantities corresponding to the exact solution, and find that
 if $n>3/2$ the expansion of our universe can be accelerated
 without invoking dark energy. Also, we test the exact
 solution of this $f(T)$ theory with the latest Union2 SNIa
 dataset which consists of 557 SNIa, and find that it can
 be well consistent with the observational data in fact.

Some remarks are in order. Firstly, although there are many
 works concerning Noether symmetry in the literature, we
 still find something new by considering Noether symmetry
 in $f(T)$ theory. For instance, we take $f(R)$ theory for
 comparison. It is well known that the equations of motion
 in $f(R)$ theory are 4th order while they are 2nd order in
 $f(T)$ theory; $R$ contains $\ddot{a}$ and $\dot{a}$, but
 $T$ contains only $\dot{a}$. Therefore, the corresponding
 point-like Lagrangian of $f(R)$ theory contains $\dot{R}$,
 but the one of $f(T)$ theory does not contain $\dot{T}$.
 Note that in the $L_{\bf X}{\cal L}=0$ equation,
 the Euler-Lagrange equation and the total energy
 (Hamiltonian) corresponding to Lagrangian,
 $\partial {\cal L}/\partial\dot{q}_i$ plays an important
 role. The absence of $\dot{T}$ in the point-like Lagrangian
 of $f(T)$ theory makes difference and brings something new.
 Secondly, unlike other theories (e.g., $f(R)$ theory, which
 has been studied for more than ten years), $f(T)$ theory
 became an active field just from 2010. Due to its relatively
 short history, many aspects of $f(T)$ theory are unexplored
 in fact. For instance, to our knowledge, the reconstruction
 of $f(T)$ from cosmological observations is absent in the
 literature so far, unlike $f(R)$ theory. We do not know what
 forms of $f(T)$ are the suitable ones, or what forms of
 $f(T)$ have a good motivation from fundamental theories and
 principles. So far, {\em all} $f(T)$ forms considered in the
 literature are written {\em by hand}. In such a situation, it
 is important to find a $f(T)$ form from some principles. Of
 course, one might argue that Noether symmetry is not really
 a physical guiding principle. But, why we deny to use
 Noether symmetry to shed some light on the unknown side of
 $f(T)$ theory? So, we consider that it is better to keep an
 open mind. Finally, we have shown in Sec.~\ref{sec5} that
 the resulting $f(T)$ theory from Noether symmetry can be well
 consistent with the observational data. We are not playing
 mathematical games with Noether symmetry. Instead, we have
 shown that the $f(T)$ motivated by Noether symmetry can be
 tested by realistic cosmological data, and then it can be
 considered as a serious theory of modified gravity and an
 alternative to dark energy.


\section*{ACKNOWLEDGEMENTS}
We thank the referee for quite useful comments and suggestions,
 which help us to improve this work. We are grateful to
 Professors Rong-Gen~Cai and Shuang~Nan~Zhang for helpful
 discussions. We also thank Minzi~Feng, as well as Hao-Yu~Qi
 and Xiao-Peng~Ma, for kind help and discussions. This work
 was supported in part by NSFC under Grants No.~11175016 and
 No.~10905005.

\renewcommand{\baselinestretch}{1.1}


\end{document}